# Multiple Antenna Technologies

Manar Mohaisen | YuPeng Wang | KyungHi Chang

The Graduate School of Information Technology and Telecommunications

INHA University

## ABSTRACT

Multiple antenna technologies have received high attention in the last few decades for their capabilities to improve the overall system performance. Multiple-input multiple-output systems include a variety of techniques capable of not only increase the reliability of the communication but also impressively boost the channel capacity. In addition, smart antenna systems can increase the link quality and lead to appreciable interference reduction.

## I. Introduction

Multiple antennas technologies proposed for communications systems have gained much attention in the last few years because of the huge gain they can introduce in the communication reliability and the channel capacity levels. Furthermore, multiple antenna systems can have a big contribution to reduce the interference both in the uplink and the downlink by employing smart antenna technology.

To increase the reliability of the communication systems, multiple antennas can be installed at the transmitter or/and at the receiver. Alamouti code is considered as the simplest transmit diversity scheme while the receive diversity includes maximum ratio, equal gain and selection combining methods. Recently, cooperative communication was deeply investigated as a mean of increasing the communication reliability by not only considering the mobile station as user but also as a base station (or relay station). The idea behind multiple antenna diversity is to supply the receiver by multiple versions of the same signal transmitted via independent channels.

On the other hand, multiple antenna systems can tremendously increase the channel capacity by sending independent signals from different transmit antennas. BLAST spatial multiplexing schemes are a good example of such category of multiple antenna technologies that boost the channel capacity.

In addition, smart antenna technique can significantly increase the data rate and improve the quality of wireless transmission, which is limited by interference, local scattering and multipath propagation. Through shaping the antenna radiation pattern and adaptively adjusting the antenna weight vector, smart antennas improve the communication link quality by increasing the received signal power and suppressing the interference.

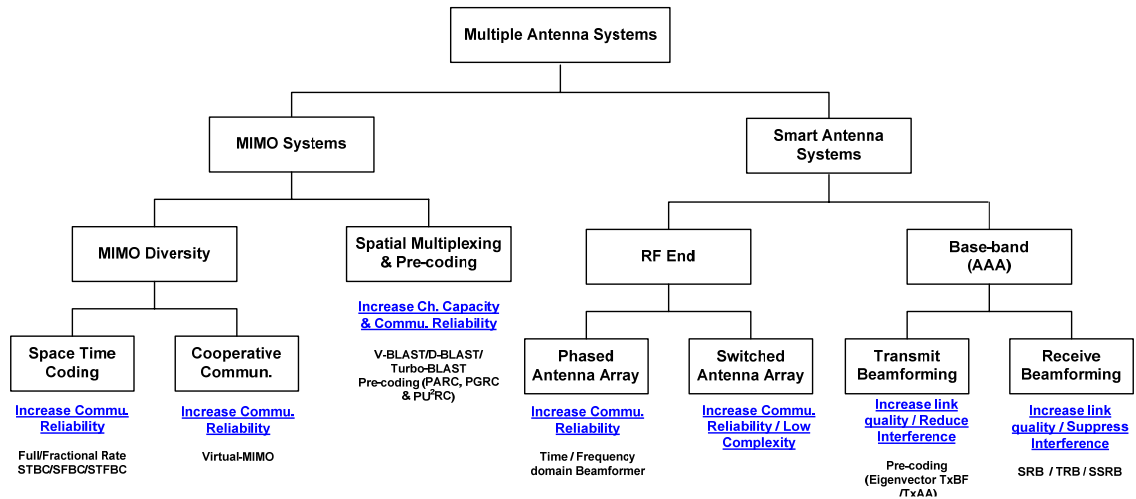

Fig. 1. Multiple antenna technologies.

Besides, on-line calibration technique is also adopted to correct the errors due to the distortions and nonlinearity of the radio frequency components in the antenna array system.

Fig. 1 summarizes the different multiple antenna technologies and gives some examples of these technologies.

This paper is organized as follows: in section II we present multiple antenna diversity schemes employed at the transmitter or/and at the receiver. Spatial multiplexing presented by BLAST schemes is detailed in section III. Section IV is dedicated to some advanced multiple input multiple-output (MIMO) systems including multi-user MIMO and cooperative communications. While techniques related to the smart antennas such as phased antenna array, switched beam antenna array, and adaptive antenna array are described in Section V. Finally, we conclude in Section VI.

## II. MIMO Diversity

In communication systems, we have to increase the reliability of the communication operation between transmitter and receiver while maintaining a high spectral efficiency. The ultimate solution relies in the use of *diversity*, which can be viewed as a form of redundancy [1]. There are many diversity techniques that can be applied to communication systems; we mention herein time diversity, frequency diversity, and spatial diversity or any combination of these three diversities. In *time diversity*, the same information-bearing signal is transmitted in different time slots where a good gain can be achieved when the duration between the two slots, in which the same symbol is transmitted, is greater than the coherence time of the channel. In *frequency diversity*, the same information-bearing signal is transmitted on different subcarriers where a good diversity gain can be achieved when the separation between subcarriers is greater than the coherence bandwidth.

Finally, in *spatial diversity*, the same information-bearing signal is transmitted or received via different antennas where the maximum gain can be achieved when the fading occurring in the channel is independent (or low correlated). In the receiver, diversity gain can be achieved by combining the redundant signals arriving via independent (or lowly correlated) channels.

Fig. 2 shows some possible combinations of transmit diversity which can be achieved when employing multiple transmit antennas.

In the following section, we present some famous space-time block codes applied at the transmitter side. We present also the combining techniques used when different versions of the information-bearing signal are received. Finally, we present a scheme that includes transmit and receive diversities.

## 2.1 Space Diversity at the Transmit Side

The basic idea of the use of transmit diversity is to reduce the mobile station (MS) receiver complexity while improving the detection performance.

The pioneering work in the transmit diversity was done by Alamouti where he proposed his famous 2×1 space-time code. Alamouti scheme achieves diversity gain while requiring only a linear decoder.

Later on, Tarokh et al. proposed a generalized theory of the complex orthogonal space-time codes. Based on Tarokh work, more than two antennas can be used and the code rate can be fractional.

In the following we present the two different types of space-time codes.

### 2.1.1 Complex Orthogonal Space-Time Codes

For this type of space time codes, the following conditions must be satisfied

- Square transmission matrix (number of transmit antennas $N_t$ equal to number of used time slots *m*)
- A unity code rate (number of used time slots *m* equals to number of transmitted symbols *l*)
- Orthogonality of the transmission matrix in the time and space domains ($SS^H = S^H S$) where $S^H$ is the conjugate transpose of $S$.

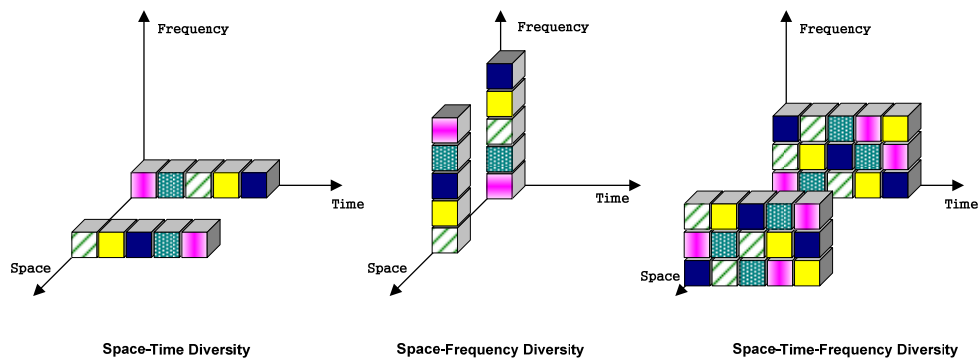

Fig. 2. Transmit diversity.

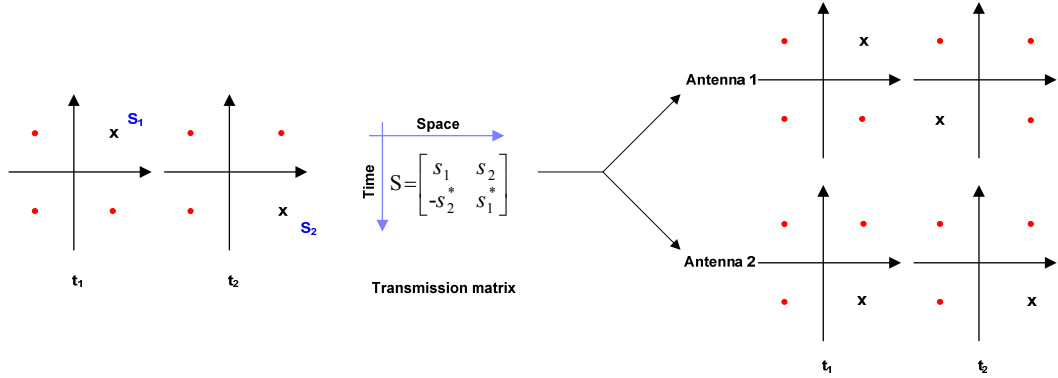

Fig. 3. Alamouti scheme example with QPSK modulation.

As said before, the simplest complex orthogonal space-time code is the Alamouti code which uses two transmit antennas and one receive antenna. Furthermore, Alamouti scheme requires that the fading channel envelope remains constant over two time slots.

Fig. 3 shows an example of the encoding process of Alamouti scheme with QPSK modulation [2], [3].

Fig. 4 shows the receiver structure used for decoding the combined received symbols.

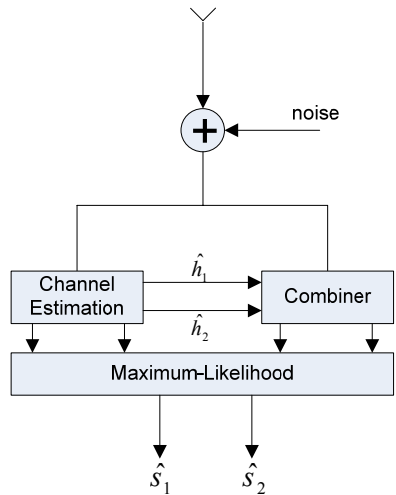

Fig. 4. Alamouti code receiver.

At the receiver the following signals are received (with applying the complex conjugate to the received signal at $t_2$)

$$\begin{bmatrix} y_1 \\ y_2^* \end{bmatrix} = \begin{bmatrix} h_1 & h_2 \\ h_2^* & -h_1^* \end{bmatrix} \begin{bmatrix} s_1 \\ s_2 \end{bmatrix} + \begin{bmatrix} n_1 \\ n_2^* \end{bmatrix} \quad (1)$$

The linear combiner multiplies the received symbols by the Hermitian transpose of the channel matrix (for simplicity, we consider that channel is perfectly estimated). The output of the linear combiner is then given by

$$\begin{bmatrix} x_1 \\ x_2 \end{bmatrix} = \begin{bmatrix} h_1^2 + h_2^2 \end{bmatrix} \begin{bmatrix} s_1 \\ s_2 \end{bmatrix} + \begin{bmatrix} w_1 \\ w_2 \end{bmatrix} \quad (2)$$

Maximum-Likelihood (ML) decoder is then applied to get the transmitted symbols. As one can see, the simplicity of the receiver is due to the spatio-temporal orthogonality of the transmission matrix.

A complex orthogonal space-time code using 4 or 8 antennas was proposed by Tarokh et al. in [4].

### 2.1.2 Generalized Complex Orthogonal Space-Time Codes

The search for space-time codes with more than two antennas was started by Tarokh, Jafarkhani, and Calderbank. Their work has built the basis for a theory of generalized complex orthogonal designs.

Generalized complex orthogonal designs are distinguished from Alamouti code by the following

- A non-square transmission matrix (number of used time slots ≠ number of Tx antennas)
- A fractional code rate (number of transmitted symbols < number of used time slots)
- Orthogonality of the transmission matrix is only guaranteed in the time sense.

As a consequence of these characteristics, the spectral efficiency is reduced and the number of time slots over which the channel should be constant is increased.

The transmission matrix of a generalized complex space-time code with 3 antennas, 4 transmitted symbols and 8 used time slots is given by [5]

$$\mathbf{G_3} = \begin{bmatrix} s_1 & s_2 & s_3 \\ -s_2 & s_1 & -s_4 \\ -s_3 & s_4 & s_1 \\ -s_4 & -s_3 & s_2 \\ s_1^* & s_2^* & s_3^* \\ -s_2^* & s_1^* & -s_4^* \\ -s_3^* & s_4^* & s_1^* \\ -s_4^* & -s_3^* & s_2^* \end{bmatrix} \quad (3)$$

In the literature, more research was done to increase the rate of the space-time codes. For more details refer to [6].

Table 1 summarizes the difference between Alamouti and space-time code characterized by the transmission matrix $\mathbf{G_3}$.

These shown coding schemes can be transmitted in the space-time domains, space-frequency domains or in space-frequency-Time domains. These coding schemes are thus known as ST, SF, and STF coding, respectively [7].

### 2.1.3 Cyclic Delay Diversity (CDD)

CDD can be considered as a very simple transmit diversity scheme. CDD can achieve transmit artificial frequency diversity by selecting appropriate transmit delays. In this method, multiuser diversity, obtained by scheduling based on frequency domain channel response, can be improved by adjusting the delay spread (at the transmitter) which is done by controlling the delay values dependent on the channel condition [8], [9].

## 2.2 Space Diversity at the Receive Side

In space diversity at the receive side, multiple antennas are used in the receiver with sufficient spacing between antennas in such a way mutual correlation between antennas is reduced and as consequence diversity gain is increased [10]. To get diversity gain at the receiver, received signals from different antennas are combined. There are four combining methods, namely, select combining (SC), maximal-ratio combining (MRC), equal-gain combining (EGC), and square-law combining. The first three schemes are linear while the last requires a non-linear receiver.

Fig. 5 shows a simplified block diagram of the linear combining schemes which differ in the weighting vector **w.** In SC, the signal at the branch with maximum signal to noise ratio (SNR) is selected and other received signals are discarded.

The weighting vector $\mathbf{w} = (w_1, w_2, \ldots, w_M)$ is the $N^{th}$ column of the identity matrix of size $M$ where the $N^{th}$ branch has the maximum SNR.

Table 1. Comparison between Alamouti and generalized complex space-time codes.

| Space-time code | Number of Tx antennas | Number of transmitted symbols, *l* | Number of used time slots, *m* | Orthogonality of Tx matrix | Rate = *l/m* |
|---|---|---|---|---|---|
| S | 2 | 2 | 2 | Spatio-temporal sense | 1 |
| $G_3$ | 3 | 4 | 8 | Only temporal sense | 1/2 |

Table 2. Comparison between diversity combining schemes.

| Scheme | Requiring CSI | Outage Probability $F(x)$ | Application |
|---|---|---|---|
| SC | No | $\left[1-e^{-x/\bar{y}_c}\right]^M$ | No constraints |
| MRC | Yes | $1-e^{-x/\bar{y}_c}\sum_{k=0}^{M-1}\frac{1}{k!}\left(\frac{x}{\bar{y}_c}\right)^k$ | No constraints |
| EGC | Yes | No closed form for M > 2 | No constraints |
| SLC | No | - | FSK or DS-CDMA |

As one can see, the SC scheme does not require any channel information except that of SNR. On the other hand, MRC and EGC schemes require the channel state information (CSI) or a part of it (channel envelope, phase, delay). MRC scheme weights the received signals according to their reliability; a more reliable signal has a high weight while a less reliable signal has a small weight. Also, the channel phase distortion is compensated. Finally, signals are aligned then combined. On the other hand, EGC scheme can be viewed as a simplified version of MRC where signals are weighted equally (i.e. the weighting vector **w** = [1, 1, …,1]$_M$) then aligned before being combined coherently. In practice, the phase at different branches can't be often estimated. So, EGC and MRC can't be employed. In such situations, square law combining (SLC) can be applied to obtain spatial diversity without requiring phase estimation. Unlike linear combining schemes, SLC scheme can only be applied to modulation schemes which preserve some sort of orthogonality including frequency-shift keying (FSK) or direct-sequence CDMA [5], [11].

Table 2 summarizes a comparison between combining schemes. It is known that, from a performance point of view, MRC is optimum and gives the best performance among the pre-described combining schemes.

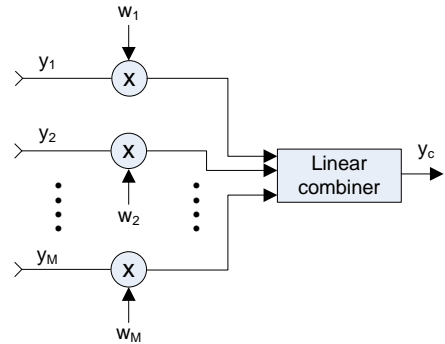

Fig. 5. Simple block diagram of linear combining schemes.

### 2.3 Combined Transmit/Receive Diversity

Spatial diversity schemes explained in the previous two sections can be combined together to achieve diversity at both receive and transmit sides. An example of such a hybrid spatial diversity scheme is the 2×2 Alamouti/MRC MIMO system [12].

## III. Spatial Multiplexing

As shown in the previous section, MIMO diversity can be used in the transmitter or the receiver sides or in both to increase the reliability of the communication. In this section we talk about spatial multiplexing schemes which are for goal to increase the channel capacity.

The most known spatial multiplexing schemes are the BLAST family which includes Vertical-BLAST, Diagonal-BLAST, and Turbo-BLAST. The acronym BLAST stands for "*Bell Laboratories Layered Space-Time*".

### 3.1 Diagonal-BLAST

D-BLAST was originally proposed by Foschini [13]. In D-BLAST, the symbols to be transmitted are arranged on the diagonals of the space-time transmission matrix where elements under the diagonal are padded with zeros. Fig. 6-a depicts the structure of the D-BLAST transmitter for four transmit antennas. At first, the bit stream is de-multiplexed into four parallel streams which are encoded and modulated independently. Encoded-modulated streams are cycled over time. Equation (4) is an example of the transmission matrix when using four transmit antennas.

$$S = \begin{bmatrix} s_{1,1} & s_{1,2} & s_{1,3} & s_{1,4} & \cdots & s_{1,K-1} & s_{1,K} & 0 & 0 & 0 \\ 0 & s_{2,1} & s_{2,3} & s_{2,4} & \cdots & s_{2,K-2} & s_{2,K-1} & s_{2,K} & 0 & 0 \\ 0 & 0 & s_{3,1} & s_{3,2} & \cdots & s_{3,K-3} & s_{3,K-2} & s_{3,K-1} & s_{3,K} & 0 \\ 0 & 0 & 0 & s_{4,1} & \cdots & s_{4,K-4} & s_{4,K-3} & s_{4,K-2} & s_{4,K-1} & s_{4,K} \end{bmatrix} \quad (4)$$

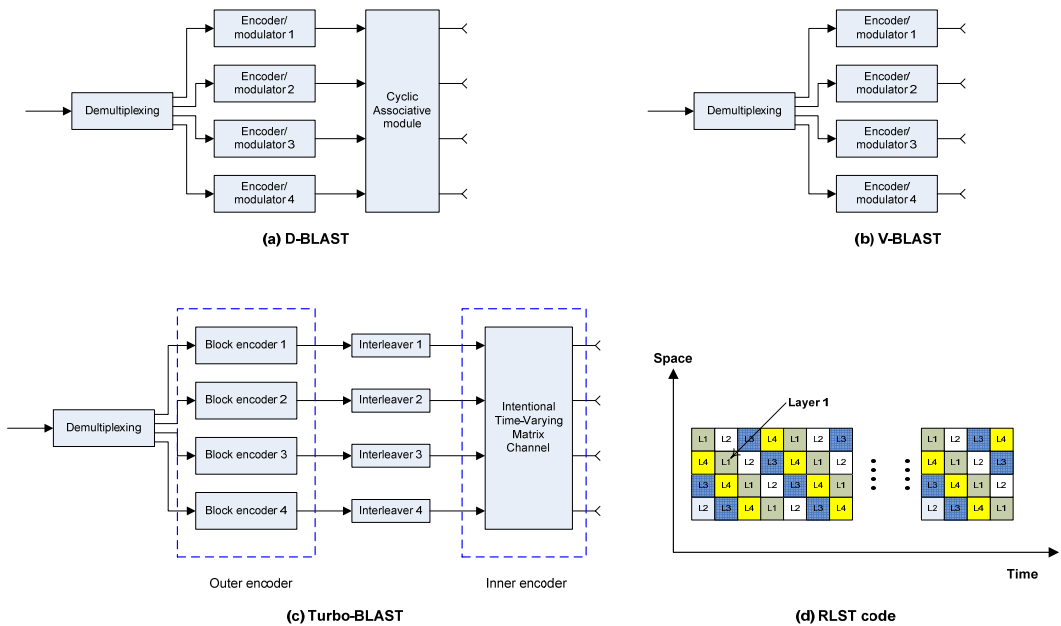

Fig. 6. Transmitter block diagrams for BLAST family using four transmit antennas.

Table 3. MIMO systems diversity orders.

| MIMO Configuration | Diversity order |
|---|---|
| STBC | $N_t \times N_r$ |
| BLAST | $N_r - N_t + 1$ |

The first diagonal of $S$ is transmitted via the first antenna; the second diagonal is transmitted via antenna 2, and so on.

## 3.2 Vertical-BLAST

A simplified version of D-BLAST was proposed by Wolniansky known as Vertical-BLAST or V-BLAST [14]. In V-BLAST, incoming data stream is de-multiplexed into $N_t$ streams each of which is encoded and modulated independently and sent on an antenna of its own. V-BLAST high-level diagram is depicted in Fig. 6-b where four antennas are used at the transmit side. Compared to D-BLAST, V-BLAST does not include cycling over time, the complexity is significantly reduced. In addition, unlike D-BLAST, V-BLAST does not include any space-time wastage. At the receiver, transmitted symbols can be decoded using *ordered serial interference-cancellation (OSIC) detector*. For the OSIC to work properly, the number of receive antennas $N_r$ must be at least as large as the number of transmit antennas.

## 3.3 Turbo-BLAST

Turbo-BLAST was first described by Sellathurai and Haykin [15]. The Turbo-BLAST transmitter structure is depicted in Fig. 6-c. The data stream bits are firstly demultiplexed into $N_t$ parallel streams which are encoded independently using the block encoder (outer encoder) (i.e. channel coding). The output streams of the outer encoder are interleaved independently and passed to the inner encoder. The mission of the outer encoder is to achieve *random-layered space-time* (RLST) coding.

The structure of the RLST encoder, with periodical cyclic space-time interleaving is depicted in Fig. 6-d. For optimal performance of the RLST code, the receiver should employ the *maximum a posteriori probability (MAP)* decoding algorithm. Nevertheless, the complexity of the MAP decoding algorithm is very high (increases exponentially with $N_t$). To decrease the complexity of the receiver, the near-optimal turbo-like receiver can be used. This near-optimal turbo-like receiver is known as *iterative detection and decoding (IDD) receiver*.

Before going further, we list in Table 3 a comparison between diversity order of the different space-time coding and the BLAST family schemes.

# IV. Advanced Topics

## 4.1 Single and Multi CodeWord MIMO

In single codeword (SCW) MIMO, an encoded packet is distributed across many streams to form the MIMO transmission. Feedback is used to control the rank of the MIMO transmission (number of streams used) as well as the overall rate of transmission. In multiple codeword MIMO, several separately encoded packets are transmitted independently over the multiple streams. Here the rate of each stream can be controlled with feedback [16] and [17].

## 4.2 Single-User MIMO and Multi-User MIMO

In single-user MIMO, already explained techniques in previous sections are used where the channel capacity grows linearly with *min(Nt, Nr)* [18].

For multi-user MIMO, which is of high interest research topic, it was shown that for $N_t$ transmitting antennas (at the base station) and $N_r$ users, the same overall capacity can be achieved. This later work was encouraged by applying dirty paper coding [19] where results showed that if the transmitter knows the interfering signal, then the channel capacity will not be affected by the presence of the interference [20]. On the other hand, multi-user MIMO can integrate beamforming to apply spatial division multiple access (SDMA).

### 4.3 Cooperative Communication and Virtual MIMO

In cooperative communication, a mobile can act as both a user and relay. As consequence, mobile sends to the base station its own data bits and some of other mobiles (sometimes called partner) information bits. Fig. 7 shows a cooperative cellular system where for simplicity we consider three cooperative users and one base station [21]. As depicted in Fig. 7, user 1 cooperates with users 2 and 3 to send its own information. As a result, the overall cooperative system can be seen as virtual-MIMO (V-MIMO) and in the above example it is $3 \times N_{BS}$ MIMO system (for the uplink) where $N_{BS}$ is the base station's number of receive antennas. Users 2 and 3 can simply amplify and forward user 1 received information or detect and forward [22]. Another method of cooperation is the coded cooperation where different coded portions are sent via different fading channels [23].

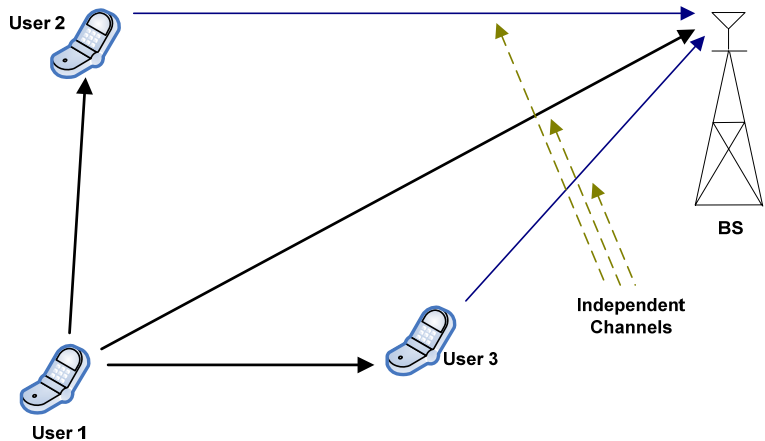

Fig. 7. Cooperative communication and virtual-MIMO.

### 4.4 Pre-Coded MIMO with Rank Adaptation

#### 4.4.1 Per Antenna Rate Control (PARC)

PARC can be considered as a closed-loop MIMO system where transmitter uses *channel quality indication* (CQI) fed by the receiver to select the best modulation and coding schemes per antenna. Fig. (8) shows a general PARC transmitter structure with 4 transmit antennas [24].

#### 4.4.2 Per Group Rate Control (PGRC)

In PARC, a CQI feedback is necessary for each transmit antenna. This increases the uplink overhead.

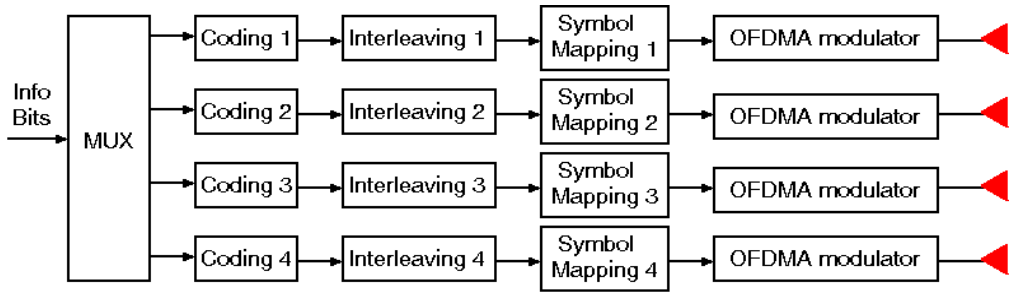

Fig. 8. Per Antenna Rate Control (PARC).

To solve this problem, PGRC is used where a feedback is required per group of antennas. This reduces the feedback information while maintaining almost the same performance of PARC [24].

### 4.4.3 Per User Unitary Rate Control (PU$^2$RC)

PU$^2$RC is a multi-user closed-loop MIMO system. Each user feeds back the CQI to the base station. The base station uses the CQI to determine the modulation and coding schemes per user. In addition, base station can apply unitary pre-coding and adaptively select the number of transmit antennas (rank adaption) [25].

## V. Smart Antennas

### 5.1 Introduction

Smart antenna was born in the early 1990s when well developed adaptive antenna arrays originate from Radar system. Later, Smart antenna technique is applied in wireless communications system. Recently, Smart antenna technique has been proposed as a promising solution to the future generations of wireless communication systems, such as the Fourth-Generation mobile communication systems, broadband wireless access networks, where a wide variety of services through reliable high-data rate wireless channels are expected. Smart antenna technique can significantly increase the data rate and improve the quality of wireless transmission, which is limited by interference, local scattering and multipath propagation [26], [27]. Smart antennas offer the following main applications in high data-rate wireless communication systems [28], [29]:

- Spatial Diversity
- Co-channel interference reduction
- Angle reuse or space division multiple access (SDMA)
- Spatial multiplexing

Smart antenna system can be categorized into three main groups: Phased antenna array system, switched beam systems, and adaptive antenna array system. To match the characteristics in each radio frequency chain of the transmitter and receiver, on-line calibration is required in smart antenna systems. On-line calibration technique can compensate the errors such as the distortions of radio frequency components due to small environment changes, the nonlinear characteristics of mixer, amplifier and attenuator, I/Q imbalance errors, etc.

### 5.2 Phased Antenna Array System

Phased antenna array is a group of antennas in which the relative phases of the respective signals feeding the antennas are varied in such a way that the effective radiation pattern of the array is reinforced in a desired direction and suppressed in undesired directions.

Phased antenna array system is usually utilized in *radio frequency* (RF) or *intermediate frequency* (IF) with the system central frequency larger than 10 GHz, such as satellite communication system [30]. There are two main different types of phased arrays, also called beamformers. There are time domain beamformers and frequency domain beamformers.

**5.3 Switched Beam System**

The switched beam method is considered as an extension of the current sectorization scheme. In the switched beam approach, the sector coverage is achieved by multiple predetermined fixed beam patterns with the greater gain placed in the centre of a beam [30]. When a mobile user is in the vicinity of a beam, then the signals at the output ports will be given as in (5). This enables the switched beam system to select the signal from the output port corresponding to that beam. As the mobile moves to the coverage of another beam during the call, the system monitors the signal strength and switches to other output ports as required. A basic switched beam antenna architecture is shown in Fig. 9. And Fig. 10 illustrates the produced antenna pattern with 4 antennas.

$$y_i(t) = s_i(t)G_i(t)\sum_{l=1}^{L} I_l(t)G_{li}(\theta_l) \quad (5)$$

where $y_i(t)$ is the total signal appearing at port $i$, $s_i(t)$ is the signal source, $I_l(t)$ is the interfering signal source located at arbitrary angles $\theta_l$, $G_i$ is the transfer function between signal source along the main beams and their corresponding output ports, $G_{li}$ is the transfer function between interference signal $l$ and port $i$.

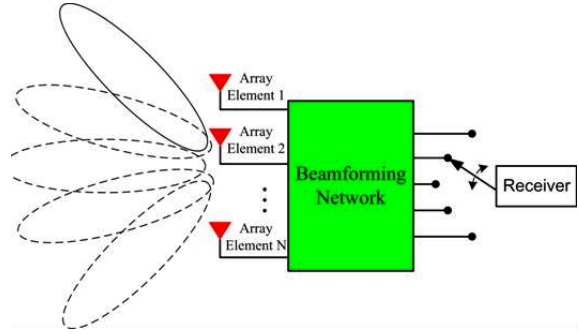

Fig.9. Functional block diagram of switched beam system.

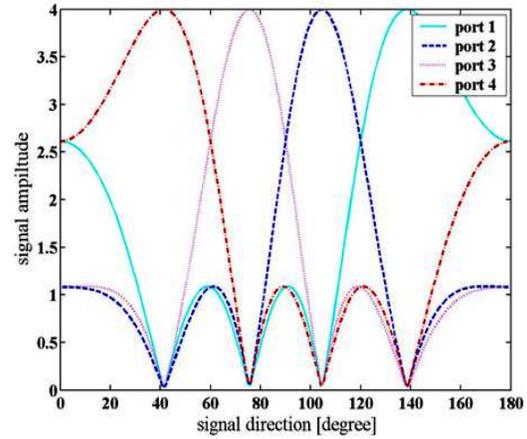

Fig.10. Produced antenna pattern of switched beam system with 4 antennas.

Switched beam systems can offer several advantages, including

- Low complexity and cost. Since switched beam system only requires a beamforming network, RF switches, and simple control logic, they are relatively easy and cheap to implement.
- Moderate interaction with base station receivers. In practice, switched beam system can simply replace conventional sector antennas without requiring significant modifications to the radio base station antenna interface or the baseband algorithms implemented at the receiver.
- Coverage extension. The antenna array aperture gain will boost the link

budget, which could be translated to a coverage extension.

## 5.4 Adaptive Antenna Array System (AAA)

Adaptive antennas date back to 1959. The original work was attributed to L. C. Van Atta's work, Electromagnetic Reflection. Since then, adaptive beamforming techniques have been employed to remove unwanted noise and jamming from the output, mainly in military applications. With the thriving commercial wireless communication industry and the advancing microprocessor technologies, the adaptive beamforming techniques have found their applications in commercial wireless communications. With powerful *digital signal processing* (DSP) hardware at the base-band, algorithms could control antenna beam patterns adaptively to the real signal environment, forming beams towards the desired signals while forming nulls to co-channel interferers. Thus, the system performance is optimized in terms of link quality and system capacity [31]. Adaptive antenna array can be utilized in the transmitter side, which is known as *transmit beamforming* (TxBF) or in the receiver side, which is called *receive beamforming* (RxBF).

### 5.4.1 Transmit Beamforming (TxBF)

The implementation of adaptive antenna array technique in a handset is difficult with today`s hardware due to its limitations in size, cost, and energy storage capability, while it is feasible to adopt antenna arrays at base stations. Transmit beamforming provides a powerful method for increasing downlink capacity [32]-[35]. The idea of TxBF is similar to the pre-coded MIMO technique but with different strategies to calculate the transmit weight vector. TxBF adjusts the antenna main lobe towards to the desired user and reduce the interference to other users. A simple illustration of TxBF is shown in Fig. 11.

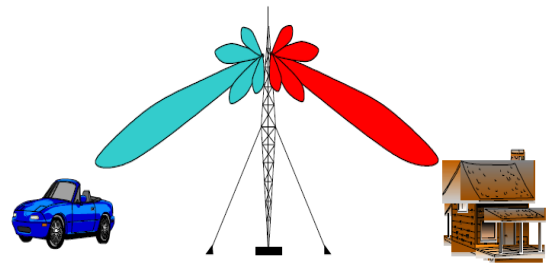

Fig.11. An illustration of TxBF.

**Eigenvector TxBF Algorithm**

Eignenvector TxBF algorithm is widely used for TxBF. The eigenvector of the spatial covariance channel matrix is calculated as

$$\mathbf{R}_{ss} = \lambda \mathbf{H} \qquad (6)$$

where $\mathbf{R}_{ss}$ is the autocovariance matrix of the desired user`s signal, and $\mathbf{H}$ is the spatial covariance channel matrix. The eigenvector $\lambda_{max}$ which corresponds to the largest eigenvalue will be selected as the weight vector [36]. One example of beam pattern for 4 uniform linear array elements is shown in Fig 12.

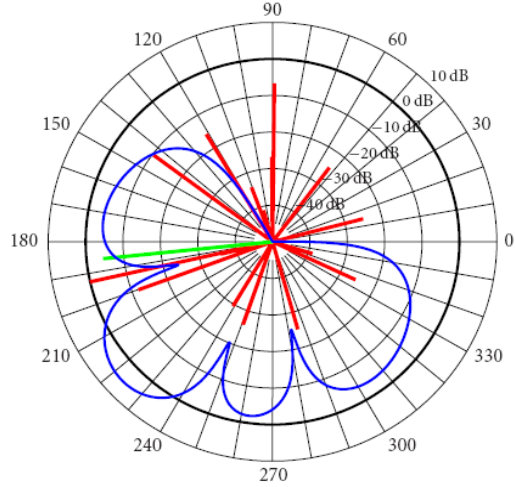

Fig. 12. Example beam pattern of 4 antenna elements in a sectorized system for a single sector (main beam direction is 240° ).

**Transmit Adaptive Array (TxAA) Algorithm**

Transmit adaptive array (TXAA) is a technique in which the user periodically sends quantized estimates of the optimal transmit weights to the BS via a feedback channel. The transmitter weights are optimized to deliver maximum power to the user. The optimal transmit weights are given by

$$\mathbf{w} = \mathbf{H}^H / \mathbf{H}\mathbf{H}^H \qquad (7)$$

where **w** is the transmit weight vector and **H** is the channel matrix.

The weights are normalized so that the total transmitted power is not altered. In the case of multipath channels emanating from each antenna, the optimal weights will be given by the principal eigenvector of the channel correlation matrix $\mathbf{H}^H\mathbf{H}$.

**5.4.2 Receive Beamforming (RxBF)**

Beamforming also can be applied in the uplink to improve the link quality and suppress the co-channel interference, which is known as *receive beamforming* (RxBF). Through RxBF, smart antenna system can receive predominantly from a desired direction (direction of the desired source) compared to some undesired directions (direction of interfering sources). This implies that the digital processing has the ability to shape the radiation pattern to adaptively steer beams in the direction of the desired signals and put nulls in the direction of the interfering signals. This enable low co-channel interference and large antenna gain to the desired signal.

Based on the reference signals adopted in the beamforming algorithms, RxBF can be classified into *spatial reference beamforming* (SRB), *temporal reference beamforming* (TRB), and *signal structure reference beamforming* (SSRB).

**Spatial Reference Beamforming (SRB)**

Spatial reference beamforming method is sometimes referred as *direction of arrival* (DoA) method. SRB estimates the direction of arrival of the signal based on the spatial reference signal, using any of the techniques like *multiple signal classification* or *estimation of signal*

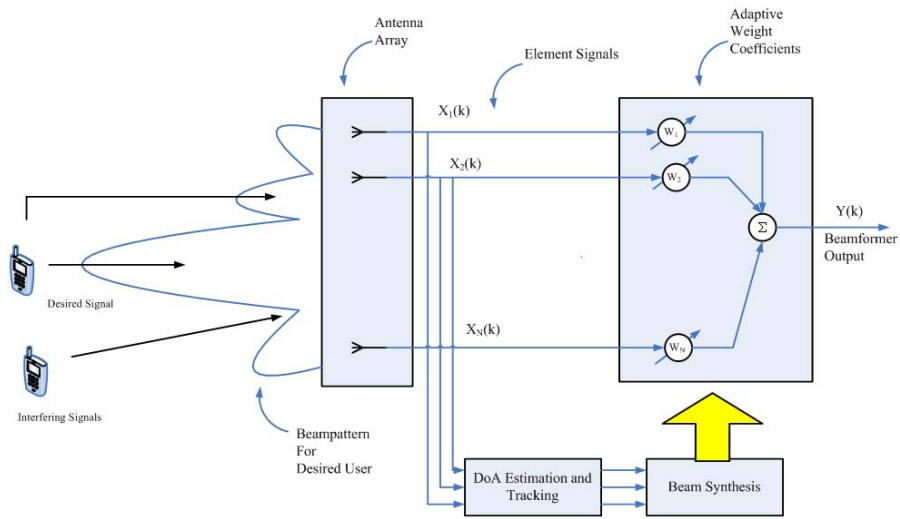

Fig. 13. A general structure of SRB.

*parameters via rotational invariance techniques* algorithms or their derivatives. They involve finding a spatial spectrum of the antenna/sensor array, and calculating the DoA from the peaks of this spectrum [37]. A general architecture of SRB algorithm is shown in Fig. 13. The general steps of SRB method are shown as follows:

- DoA Estimation
    - Arbitrary Array: MUSIC, etc.
    - Linear Array: ESPRIIT, etc.
- Beam Synthesis
    - Gram-Schmidt, etc.
- Combining
    - EGC, MRC, Wiener Filter, etc.

*Multiple signal classification* (MUSIC) algorithm estimates the DoA of the desired signal by using an eigen-space method based on a spatial reference signal. MUSIC requires intensive calculation of eigenvalues and eigenvectors of an autocorrelation matrix of the input vectors from the receiving antenna array. A general step of MUSIC algorithm is shown below:

- Collect received samples and estimate the covariance matrix of the received samples.
- Perform eigen-decomposition of the covariance matrix.
- Calculate spatial spectrum.
- Estimate DoA by locating peaks in the spectrum.

*Estimation of signal parameters via rotational invariance techniques* (ESPRIT) is also well known for the SRB method. In addition, ESPRIT has many important advantages over MUSIC algorithm [38]:

- No knowledge of the array geometry and element characteristics are required.
- Much less complex on computation.
- No calibration of the array is required.
- The algorithm simultaneously estimates the number of sources and DoA`s

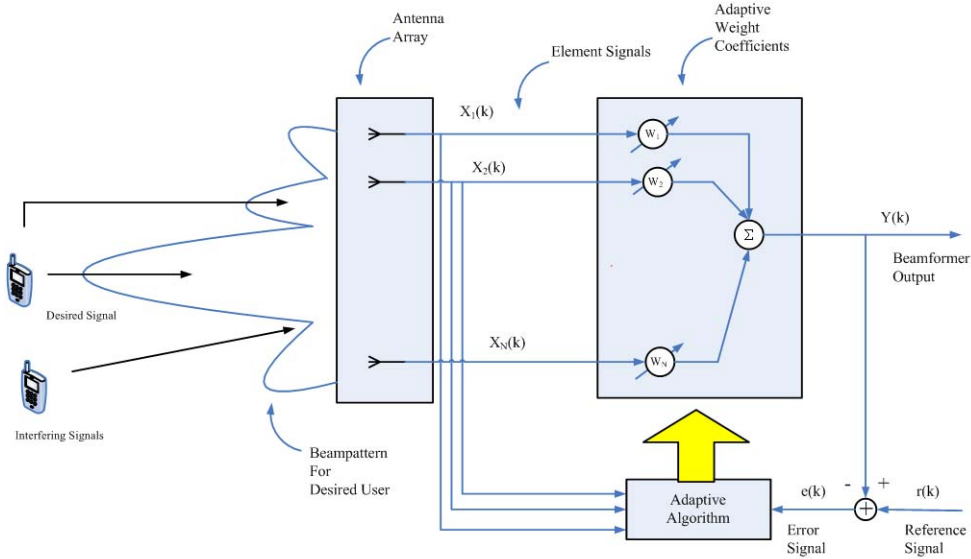

Fig. 14. A general structure of TRB.

## Temporal Reference Beamforming (TRB)

Temporal Reference Beamforming shown in Fig. 14, is a method used to create the radiation patter of the antenna array by adding constructively the phases of the signals in the DoA of the desired user, and nulling the pattern of the interfering users based on the temporal reference signal [39]. Based on the temporal reference signal and some predefined adaptive weight calculation criterion, some adaptive algorithms such as LMS (Least Mean Square), RLS (Recursive Least Squares), and DSMI (Direct Sample Matrix Inverse) algorithms, are used to adjust the weight vector of the antenna array to improve the link quality. The general characteristics of TRB are as follows:

- Good performance in multipath channel environment
- Computationally inexpensive
- Requiring Training sequence
- Difficult to apply TxBF because of the absence of DoA information

## Criterion of Adaptive Weight Calculation

In the *minimum mean-square error* (MMSE) criterion, the weights are chosen to minimize the *mean-square error* (MSE) between the beamformer output and the temporal reference signal. While in the *maximum signal-to-interference ratio* (MSIR) criterion, the weights are chosen to directly maximize the signal-to-interference ratio (SIR). And The *minimum variance* (MV) criterion chooses the weights that minimize the variance of the output power. All the above three criterions has the same form of

$$\mathbf{w}_{opt} = \beta \mathbf{R}_i^{-1} \mathbf{v} \qquad (8)$$

where $\mathbf{R}_i^{-1}$ is the inverse of the covariance matrix of the interference signals received in the antenna array and $\mathbf{V}$ is the antenna array propagation vector [40].

Let us assume that *d(t)* is the transmitted temporal reference signal and $\mathbf{R}_u$ is the covariance matrix of interference signals at the output of the beamformer. The calculation of *β* for MMSE, MSIR

Table 4. $\beta$ Calculation.

| Criterion | MMSE | MSIR | MV |
|---|---|---|---|
| $\beta$ | $\dfrac{E\{d^2(t)\}}{1+E\{d^2(t)\}\mathbf{v}^H\mathbf{R}_i^{-1}\mathbf{v}}$ | $\dfrac{E\{d^2(t)\}}{SIR}\mathbf{v}^H\mathbf{w}_{opt}$ | $\dfrac{g}{\mathbf{v}^H\mathbf{R}_u^{-1}\mathbf{v}}$ |

and MV criteria are summarized in Table 4.

**Adaptive Beamforming Algorithm**

The *least mean square* (LMS) algorithm uses the temporal reference signal to update the weights at each iteration. In the LMS algorithm, we are searching for the optimal weight that would make the array output either equal or as close as possible to the reference signal, which is the weight that minimizes the MSE. Since the MSE has a quadratic form, moving the weights in the negative direction of the gradient of the MSE should lead us to the minimum of the error surface. The weight update equation is shown in (9) [30].

$$\mathbf{w}(t+1) = \mathbf{w}(t) - \mu\mathbf{x}(t+1)\varepsilon^* \quad (9)$$

where $\mu$ is a constant, called the step size, which determines how close the weights approach the optimum value after each iteration and it controls the convergence speed of the algorithm. And $\varepsilon$ is the error signal between the temporal reference signal and the received signal at the beamformer output. $\mathbf{x}(t+1)$ is the received signal vector at the antenna array at time $t+1$.

The main drawback of the LMS algorithm is that it is sensitive to the scaling of its input. This makes it very hard (if not impossible) to choose a step size μ that guarantees stability of the algorithm. The *normalized least mean square* (NLMS) algorithm is a variant of the LMS algorithm that solves this problem by normalizing with the power of the input. The weights updating function of NLMS algorithm is shown as

$$\mathbf{w}(t+1) = \mathbf{w}(t) + \dfrac{\hat{\mu}}{a+\|x(t)\|^2}\mathbf{x}(t)\varepsilon^* \quad (10)$$

*Recursive least square* (RLS) algorithm is derived to overcome the drawback of slow convergence speed in the LMS algorithm, when the eigenvalue spread of the correlation matrix R of received signal vector **x** is large. RLS algorithm replaces the step size $\mu$ with the inverse of R. The weights are then updated using (11).

$$\mathbf{w}(t+1) = \mathbf{w}(t) - \mathbf{R}^{-1}\mathbf{x}(t+1)\varepsilon^* \quad (11)$$

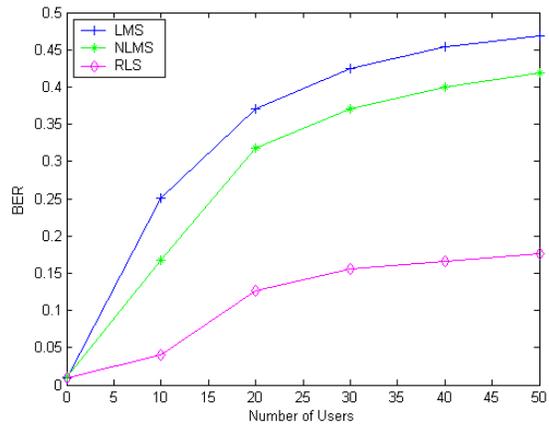

Fig. 15. Performance comparison among LMS, NLMS, and RLS algorithms.

Fig. 15 shows a simple performance comparison of the above three algorithms in OFDMA system under the Rayleigh fading channel with 8 antenna elements [41]. From this figure, we see that RLS algorithm performs best due to its faster convergence speed than LMS and NLMS algorithms.

**Signal Structure Reference Beamforming (SSRB)**

SSRB method is based on inherent structure of the transmit signal of the implicit kind reference signal. Algorithms such as blind beamforming, least squares, and constant modulus algorithms, are based on the SSRB method. SSRB method is robust against different propagation conditions and does not require the array manifold knowledge. But the convergence problem becomes the main drawback of the SSRB method.

# VI. Conclusions

In this paper we introduced the multi antenna technologies which can be considered as one of the most vivid area of research. Multiple antenna technologies were categorized into two main groups where in the first group we introduced some techniques related to spatial diversity and spatial multiplexing by outlining the gain achieved by these schemes. Furthermore, we introduced the smart antenna techniques and the up-to-date research progress in this field.

The advantages of multiple antenna systems make of them a very strong candidate to increase link reliability, increase channel capacity and reduce interference in both uplink and downlink.

# References

[1] A. Goldsmith, *Wireless Communications*, Cambridge Univ. Press, 2005.

[2] M. Alamouti, "A simple transmit diversity technique for wireless communications", *IEEE J. Sel. Area Commun.,* 16, pp. 1451-1458, 1998.

[3] G. Tsoulos, *MIMO System Technology for Wireless Communications*, Taylor and Francis, 2006.

[4] V. Tarokh et al., "Space-time block codes from orthogonal designs", *IEEE Trans. Inf. Theory,* vol. 45, no. 5, pp. 1456-1467, 1999.

[5] S. Haykin and M. Mohr, *Modern Wireless Communications,* Pearson Prentice Hall, 2005.

[6] W. Su and X-G. Xi, "Two generalized complex orthogonal space-time block codes of rates 7/11 and 3/5 for 5 and 6 transmit antennas", *IEEE Trans. Inf. Theory,* vol. 49, no. 1, pp. 313-316, Jan 2003.

[7] K. Suto, and T. Ohtsuki, "Space-time-frequency block codes over frequency selective fading channels," *IEICE Trans. Commun.,* vol. E87-B, no. 7, pp. 1939-45, July 2004.

[8] NTT DoCoMo, "Multi-degree cyclic delay diversity with frequency-domain channel dependent scheduling", *3GPP TSG-RAN WG1 meeting #44bis*, R1-060991, Mar. 2003.

[9] Samsung, "Adaptive cyclic delay diversity," *3GPP TSG-RAN1 43#*, R1-051354, 7$^{th}$ – 11st, Nov. 2005, Seoul, Korea.

[10] G. Stüber, *Principles of Mobile Comm-unication,* Kluwer, 2001.

[11] M. K. Simon and M-S. Alouini, *Digital Communication over Fading Channels*, Wiley, 2005.

[12] E. Biglieri et al., "Diversity, interference cancellation and spatial

multiplexing in MIMO mobile WiMAX systems", *to appear in IEEE WiMAX '07*, 2007.

[13] G. Foschini, "Layered space-time architecture for wireless communication in fading environment when using multi-element antennas", *Bell Labs Technical Journal,* Autumn 1996.

[14] P. Wolniansky et al., "V-BLAST: an architecture for realizing very high data rates over the rich-scattering wireless channel", *URSI International Symposium on Signals,* Systems and Electronics, 1998.

[15] M. Sellathurai and S. Haykin, "TURBO-BLAST for wireless communications: theory and experiments", *IEEE Trans. Signal Processing,* vol. 50, no., 10 pp. 2538-2546, Oct. 2002.

[16] A. Hottinen et al., "Industrial embrace of smart antennas MIMO," *IEEE Wireless Commun. Mag.*, Aug. 2006.

[17] A. Jette et al. "UMBFDD candidate proposal for IEEE 802.20," IEEE C802.20-07/09, Mar. 2007.

[18] A. Goldsmith et al., "Capacity limits of MIMO channels," *IEEE J. on Sel. Areas in Commun.*, vol. 21, no. 5, pp. 684-702, Jun. 2003.

[19] M. Costa, "Writing on dirty paper," *IEEE Trans. Info. Theory*, vol. 29, no. 3, pp. 439-441, May 1983.

[20] Q. Spencer et al., "An introduction to the multi-user MIMO downlink," *IEEE Commun. Mag.,* pp. 60-67, Oct. 2004.

[21] A. Nosratinia and A. Hedayat, "Cooperative communication in wireless networks," *IEEE Communications Magazine,* pp. 74-80, Oct. 2004.

[22] J. Laneman et al., "An efficient protocol for realizing cooperative diversity in wireless networks," *Proc. IEEE ISIT,* June 2001, pp.294.

[23] T. Hunter and A. Nosratinia, "Diversity through coded cooperation," *IEEE Trans. on Wireless Commun.,* vol. 5, no. 2, pp. 1-7, Feb. 2006.

[24] Texas Instruments, "MIMO OFDMA E-UTRA proposal for different antenna configurations," *3GPP TSG RAN1 WG1 #43*, R1-051315, Nov. 2005.

[25] Samsung, "Downlink MIMO for EUTRA," *3GPP TSG RAN1 WG1 #43*, R1-051353, Nov. 2005.

[26] S. Ohmori, Y. Yamao, and N. Nakajima, "The future generationsof mobile communications based on broadband access technologies," *IEEE Commun. Mag.,* vol. 38, pp. 134–142, Dec. 2000.

[27] K. Sheikh, D. Gesbert, D. Gore, and A. Paulraj, "Smart antennas for broadband wireless access networks," *IEEE Commun. Mag.,* vol. 37, pp. 134–142, Nov. 1999.

[28] A. Lozano, F.R. Farrokhi, and R.A. Valenzuela, "Lifting the limits on high-speed wireless data access using antenna arrays," *IEEE Commun. Mag.,* pp. 156–162, Sept. 2001.

[29] Jeffrey H. Reed," Smart antennas: A system level overview for software defined radios for creating an API", SDRF-04-I-0057-V0.00, Software Defined Radio Forum, Jan. 2004.

[30] Ahmed EI Zooghby, *Smart Antenna Engineering*, Artech House, 2005.

[31] Michael Chryssomallis, "Smart antennas", *IEEE Antennas and Propagation Mag.,* vol. 42, pp.129-136, , June 2000.

[32] H. Boche and M. Schubert, "Theoretical and experimental comparison of optimization criteria for downlink beamforming," *European*

*Trans. on Telecomm.,* vol. 12, no. 5, pp. 417–426, 2001.

[33] R. M. Buehrer, A. G. Kogiantis, S.-C. Liu, J. Tsai, and D. Uptegrove, "Intelligent antennas for wireless communications -uplink," *Bell Labs Technical Journal,* vol. 4, no. 3, pp. 73–103, 1999.

[34] H. Holma and A. Toskala, *WCDMA for UMTS*, John Wiley & Sons, 2000.

[35] A. Yener, R. D. Yates, and S. Ulukus, "Interference management for CDMA systems through power control, multiuser detection, and beamforming," *IEEE Trans. on Comm.,* vol. 49, no. 7, pp. 1227–1239, 2001.

[36] S.J. Ko, J. Heo, and K.H. Chang, "An effective downlink resource allocation for supporting heterogeneous traffic data in an OFDM/SDMA-based cellular system," *in Proc. IEEE GLOBECOM*, Nov. 2006, WLC27-5.

[37] Lal C. Godara, "Application of antenna arrays to mobile communications, part II: beam-forming and direction-of-arrival considerations", *IEEE Proc.,* vol .85, pp.1195-1245, Aug. 1997.

[38] Haardt, M., Nossek, J.A., "Unitary ESPRIT: how to obtain increased estimation accuracy with a reduced computational burden," *IEEE Trans. on Signal Processing,* vol.43, pp. 1232 -1242, 1995.

[39] B. L. P. Cheung, Simulation of Adaptive Array Algorithms for OFDM and Adaptive Vector OFDM Systems, Master of Science in Electrical Engineering of Virginia Polytech., Sept. 2002.

[40] John Litva and Titus Kwok-Yeung Lo, *Beamforming in Wireless Communication*, Artech House, 1996.

[41] J. Heo and K.H. Chang, "Transmit and receive beamforming for OFDMA/TDD system," *in Proc. ISAP*, Aug. 2005, pp.27-30.

**Manar Mohaisen**

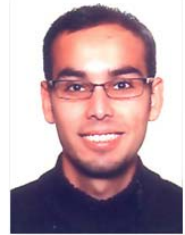

**July, 2001:** BS, Communication and Control, University of Gaza, Gaza, Palestine

**Sep., 2005**: MS， The School Polytechnic of Nice University, Sophia-Antipolis, France

**Feb., 2006 ~ Present:** Ph.D. student, The Graduate School of IT & T, INHA University

**2001 ~ 2003:** The Palestinian Telecommunication Company (JAWWAL)

<Rresearch Interests> MIMO Detection and Co-Channel Interference Cancellation

**Yupeng Wang**

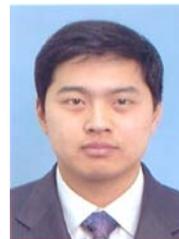

**July, 2004:** BS, Communication Engineering, Northeastern University, Shenyang, China

**July, 2006:** MS, The Graduate School of IT & T, INHA University

**Sep., 2006 ~ Present**: Ph.D. student, The Graduate School of IT & T, INHA University

<Research Interests> 3GPP LTE Systems, Radio Resource Management, MIMO Techniques, and UWB

**KyungHi Chang**

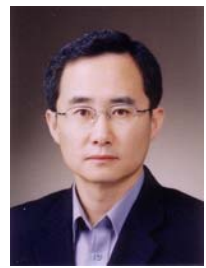

**Feb., 1985**: BS, Electronics Engineering, Yonsei University

**Feb., 1987**: MS, Electronics Engineering, Yonsei University

**Aug., 1992:** Ph.D., EE Dept., Texas A&M Univ.

**1989 ~ 1990:** A Member of Research Staff, Samsung Advanced Institute of Technology (SAIT)

**1992 ~ 2003:** A Principal Member of Technical Staff (Team Leader), Electronics and Telecommunications Research Institute (ETRI)

**2003 ~ Present:** Associate Professor, The Graduate School of IT & T, INHA University

**<Research Interests>** RTT design for IMT-Advanced & 3GPP LTE Systems, WMAN System Design, Cognitive Radio, Cross-layer Design, and Cooperative Relaying System